\documentclass[modern]{aastex62}
\usepackage{amsmath}    
\newcommand\msun {M_\odot}

\newcommand\mearth {{M_\oplus}}

\newcommand\ltsima{$\; \buildrel <\over\sim \;$}
\newcommand\simlt{\lower.5ex\hbox{\ltsima}}
\newcommand\gtsima{$\; \buildrel >\over\sim \;$}
\newcommand\simgt{\lower.5ex\hbox{\gtsima}}
\newcommand\Msi{\mathfrak{M}}

\shorttitle{No Planet Desert in RV Data}
\shortauthors{Bennett}

\begin{document}

\title{No Sub-Saturn-mass Planet Desert in the CORALIE/HARPS Radial-velocity Sample}

\correspondingauthor{David P.\ Bennett}
\email{bennettd@umd.edu}

\author[0000-0001-8043-8413]{David P.\ Bennett}
\affiliation{Laboratory for Exoplanets and Stellar Astrophysics, NASA Goddard Space Flight Center, Greenbelt, MD 20771, USA}
\affiliation{Department of Astronomy, University of Maryland, College Park, MD 20742, USA}

\author{Cl\'ement Ranc}
\affiliation{Sorbonne Universit\'e, CNRS, UMR 7095, Institut d'Astrophysique de Paris, 98 bis bd Arago, 75014 Paris, France}

\author{Rachel B.\ Fernandes}
\affiliation{Lunar and Planetary Laboratory, The University of Arizona, Tucson, AZ 85721, USA}
\affiliation{Earths in Other Solar Systems Team, NASA Nexus for Exoplanet System Science}




\begin{abstract}

We analyze the CORALIE/HARPS sample of exoplanets \citep{mayor11} found by the Doppler radial velocity 
method for signs of the predicted  gap or ``desert" at 10-$100\mearth$ caused by runaway gas accretion 
at semimajor axes of $< 3\,$AU. 
We find that these data are not consistent with this prediction. This result is similar to the finding by the MOA
gravitational microlensing survey that found no desert in the exoplanet distribution
for exoplanets in slightly longer period orbits and somewhat lower host masses
\citep{suzuki18}. Together, these results imply that the runaway  gas accretion scenario of the core 
accretion theory does not have a large influence on the final mass and semimajor axis distribution
of exoplanets.

\end{abstract}



\section{Introduction} \label{sec:intro}

The runaway gas accretion process has long been considered to be a major step in the formation
of gas giant planets in the core accretion theory \citep{pollack96}. Several authors \citep{idalin04,mordasini09a,mordasini09b}
have predicted that this runaway gas accretion process should produce a gap or  ``desert" in the 
mass distribution of planets. The abstract of  \citet{idalin04} states ``Since planets' masses grow rapidly 
from 10 to $100 M_\earth$, the gas giant planets rarely form with asymptotic masses in this 
intermediate range\rlap." They specifically indicate that this desert should exist at orbital major axes
of 0.2--$3\,$AU. The Bern population synthesis group
\citep{mordasini09a,mordasini09b} predict a lower amplitude desert, a factor of 2--3  at
masses between 30 to $100 M_\earth$. The paper presenting the data that we analyze in this paper
\citep{mayor11} claims that these data tend to confirm this prediction. They state:
``After correction of detection biases, we see even more clearly the importance of the population 
of low-mass planets on tight orbits, with a sharp decrease of the distribution between a few Earth masses and
$\sim 40\mearth$. We note that the planet population synthesis models by \citet{mordasini09a,mordasini09b} 
predicted such a minimum in the mass-distribution at precisely this mass range\rlap." 
As we shall see, our interpretation differs somewhat
from this, and it strongly contradicts a more recent version of the Bern group's population 
synthesis code \citep{emsenhuber_ngpps1,emsenhuber_ngpps1}, which predicts a deeper desert, with a planet occurrence
rate drop of a factor of $\sim 10$.

In this paper, we present the a re-analysis of the \citet{mayor11} Doppler radial velocity detected planet sample,
that was recently analyzed by \citet{fernandes19}. While \citet{fernandes19} focused primarily on the orbital
period dependence of the planet distribution, our analysis is focused primarily on the dependence on planetary mass,
so that we can investigate the proposed sub-Saturn-mass planetary desert. Before considering the \citet{mayor11} 
sample, however, we review recent results from gravitational microlensing that seem to contradict this 
cold sub-Saturn-mass desert prediction in Section~\ref{sec:snow_line_desert}. Then,
in Section~\ref{sec:mayor11}, we present the
binned, completeness corrected planet distribution with an estimate of the error bars for each bin, 
and we show that this data shows no significant 
sign of a sub-Saturn-mass planet desert. In Section~\ref{sec:forward}, we present a more powerful
forward modeling method to search for evidence of a planet desert in the \citet{mayor11} data
using a planet occurrence model that employs a Gaussian desert feature, and
again, we find no evidence of a sub-Saturn-mass planet desert. We also compare our results to the \citet{fernandes19}
analysis and recover their main result. We then consider the latest generation of planet population synthesis models
from the Bern group \citep{emsenhuber_ngpps1,emsenhuber_ngpps2} in Section~\ref{sec:bern}, and we find that our forward 
Gaussian desert model does a reasonable job at characterizing the planet desert feature in their
simulated planet population that is contradicted by the \citet{mayor11} CORALIE/HARPS data. 
\citet{emsenhuber_v_mayor} have also pointed out that deep planet desert at $20\mearth < M < 200\mearth$
seen in their models is contradicted by the
\citet{mayor11} data. Next, we show that a possible sub-Jupiter ``radius desert" that may be seen in the Kepler data
for wider orbit planets is likely to be due to the planet mass-radius relation instead of a desert in the mass distribution.
In Section~\ref{sec:conclude}, we present our conclusions
and discuss our expectations for future developments in the study of wide orbit planet demographics.

Note that the planet desert that we investigate in this paper differs from other ``deserts" that have 
been observed in or suggested by Kepler data. These are considered in Appendix~\ref{sec:kepler}.

\section{Search for a Planet Desert Beyond the Snow Line} \label{sec:snow_line_desert}

A recent comparison of theoretical population synthesis calculations, based on the core accretion theory, with
the distribution of planets detected by gravitational microlensing \citep{suzuki18}, orbiting beyond the snow line,
found that the theoretical models predicted a strong desert at mass ratios of $10^{-4} \leq q \leq 4\times 10^{-4}$
that was not reproduced by the microlensing results. Instead, the exoplanet mass ratio distribution found by 
microlensing indicated a smooth, power-law distribution down to a peak at a mass ratio of $q\sim 10^{-4}$,
with fewer planets at lower mass ratios \citep{udalski18,jung19}. This discrepancy between theory and observation
was thought to be likely due to deficiencies in the assumptions and approximations in the population synthesis 
calculations instead of problems with the fundamental concepts of the core accretion theory. One possibility that has
been suggested for our own Solar System is that captured planetesimals could heat the accreting gaseous envelope,
which would act to slow accretion, and other possibilities are that the scale height or viscosity of the protoplanetary
disk may be smaller than the values adopted for these population synthesis calculations
\citep{dobbsdisxon07,fung14,garaud07}. A reduction of the assumed disk viscosity in the Bern group's population synthesis 
model did indicate that this sub-Saturn-mass desert could disappear with lower viscosity \citep{suzuki18}, as
earlier work had indicated \citep{szulagyi14}.
Annular structures in protoplanetary disks observed by ALMA are thought by many to be due to planets in the process of
formation. A comparison of these structures in a number of protoplanetary disks also indicates that 
the growth of gas giant planets is slower than predicted by standard runaway gas accretion models \citep{nayaskshin19},
so runaway gas accretion is disfavored unless these annular structures are caused by another mechanism.

\section{Planet Occurrence for the CORALIE/HARPS Sample} \label{sec:mayor11}

\begin{figure}
\label{fig:sensitivity}
\begin{center}
\includegraphics[width=150mm]{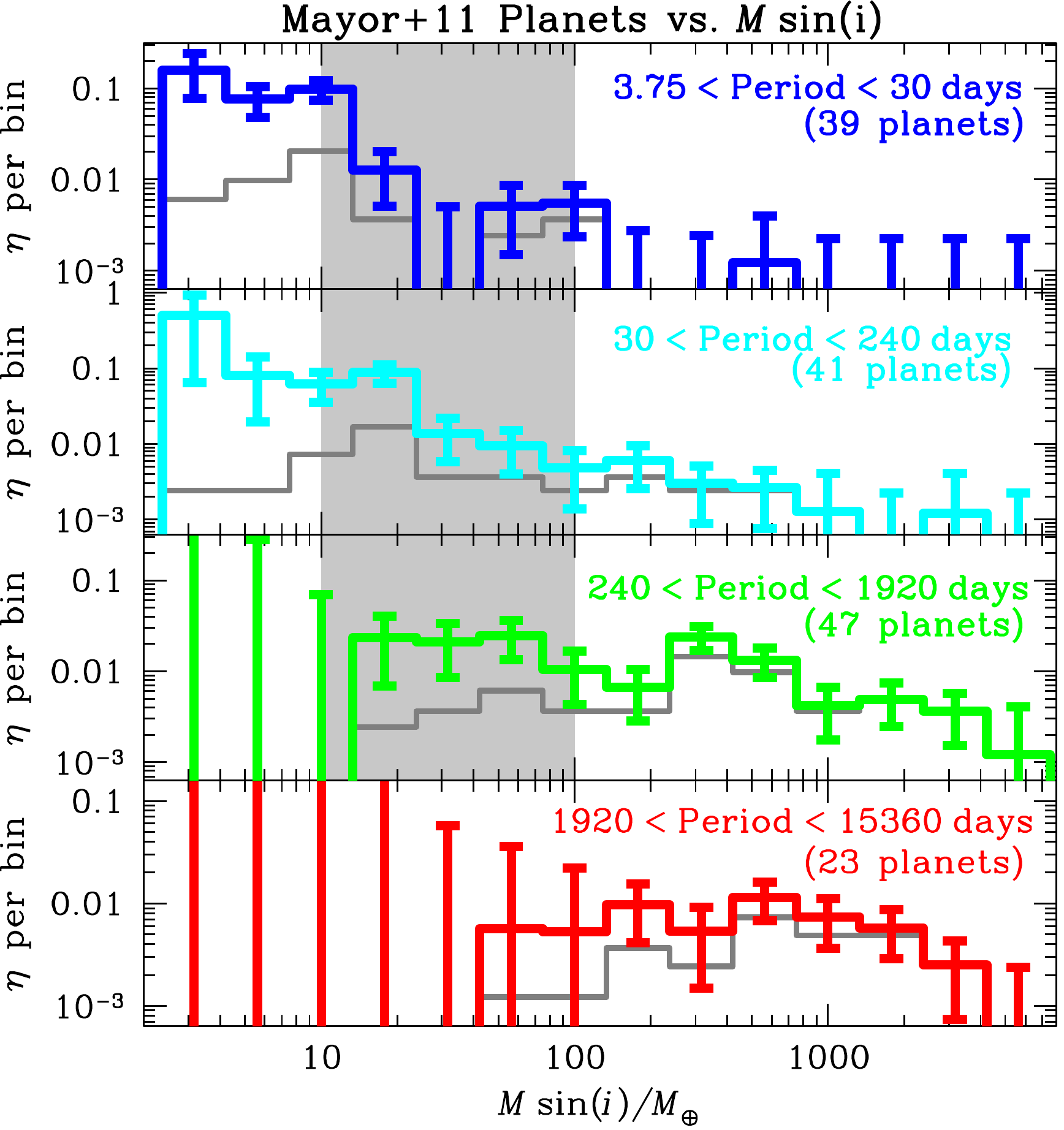}
 \end{center}
\caption{
Planet occurrence rate histograms for the planets in the \citet{mayor11} sample
in four different orbital period ranges, each spanning a factor of eight in period. The
histograms without any completeness corrections
are shown in grey and the light grey shaded regions show the
location of the ``planet desert" predicted by \citet{idalin04}.
\label{fig:mayor_hist}
}
\end{figure}

The CORALIE/HARPS data set \citep{mayor11} that we analyze in this paper was recently
analyzed by \citet{fernandes19}, who considered the distribution of planets in the 30--$6000\mearth$
mass range. They fit this with a broken power-law model that has a break and peak at
a period of $1720\pm 430\,$days. When they add a power-law for the mass distribution of 
planets in this mass range, the range of possible locations of the peak expands to $2100\pm 1200\,$days.

The \citet{fernandes19} analysis of the \citet{mayor11} data set included a tabulation of the
survey completeness from Figure 6 of \citet{mayor11}, and we use this tabulation of the
survey completeness for our analysis. As a first step in our analysis, we present the binned
$M \sin i$ distribution of the \citet{mayor11} sample in Figure~\ref{fig:mayor_hist}, where 
$M$ is the planet mass and $i$ is the inclination of the planet's orbital plane to the line-of-sight.
We break up the \citet{mayor11} sample into 4 different period ranges, each spanning a factor
of eight in period: 3.75--$30\,$days, 30--$240\,$days, 240--$1920\,$days and 1920--$15260\,$days. 
The boundary between the third and fourth period ranges is set to be $1920\,$days because this corresponds
to the approximate outer semi-major axis boundary of $3\,$AU of the \citet{idalin04} prediction for a $1\msun$
host star and is within $0.5\sigma$ of the break in the giant planet occurrence rate found by
\citet{fernandes19}.

In order to relate the planets found in the CORALIE/HARPS data set to the intrinsic planetary occurrence
rate, we correct for the survey completeness, which gives the colored histograms with error bars
shown in Figure~\ref{fig:mayor_hist}. The occurrence rate is given by
\begin{equation}
\eta = {1\over N_\star} \sum_j^{n_p} {1\over C(M_j\sin i_j,P_j)} \ ,
\label{eq-occurrence}
\end{equation}
where $N_\star = 822$ is the number of stars in the CORALIE/HARPS sample, $n_p$ is the number of 
detected planets, and $M_j$, $i_j$, and $P_j$ are the mass, orbital inclination, and the period of the
$j$th planet in the sample. The survey completeness for mass $M$, inclination $i$ and period $P$ is
given by $C(M\sin i,P)$.

The grey histograms without error bars in Figure~\ref{fig:mayor_hist}. indicate the raw, 
uncorrected histograms. The grey and colored histograms merge for large $M\sin i$ and small
periods, as expected, because these planets can be detected with nearly 100\% completeness.

In order to present the planet occurrence rate as a function of mass (or $M\sin i$) and period, we 
bin the data in both $M\sin i$ and period. 
Because the CORALIE/HARPS data set is sensitive to a very wide range of planetary orbital periods and 
masses, it is necessary to select bins that span a relatively large range in both $M\sin i$ and period in order
to have a reasonable number of planets in each bin. This complicates the issue of determining the error bars
for the completeness corrected occurrence rate for each bin, because the completeness can vary significantly
across some of the bins. Using the average completeness over the range of parameters for each bin
could induce an error if the planet occurrence rate also varies significantly over the bin size. Therefore, we
use the completeness for the $M\sin i$ and $P$ for each planet in equation~\ref{eq-occurrence}. This means 
that planets in the same bin will have different contributions to the planet occurrence rate, $\eta$, so that 
we cannot use Poisson statistics to calculate the error bars for $\eta$. Therefore, we employ a bootstrap
method to determine the error bars. We take the number of planets in each bin as the expected number 
for a Poisson distribution, and then we randomly select the planets from the list of planets in the bin.
We repeat this procedure 100,000 times for each bin to determine the root-mean-square dispersion of the 
completeness corrected $\eta$ value, which we use as the error estimate for the $\eta$ value for each
bin. For bins without any planets, we calculate the mean completeness for the bin, and then use the 
one-sided 1--$\sigma$ upper limit from Poisson statistics for zero detections to determine the 1--$\sigma$ upper limit.

The grey shaded regions in the upper three panels of Figure~\ref{fig:mayor_hist} indicate the mass range
of the planet desert predicted by \citet{idalin04}. As noted by \citet{mayor11}, there is a significant
decrease in the exoplanet occurrence rate, $\eta$, between 10 and $\sim 40\mearth$ at the shorter
orbital periods, particularly in the $3.75 < P < 30\,$days range. (With our binning, this minimum
appears at $\sim 30\mearth$.) So, planets at the bottom edge of the predicted ``desert" are 
clearly more common, but to claim a desert, the exoplanet occurrence must increase at higher masses,
and such a feature is not seen in Figure~\ref{fig:mayor_hist}.
Also, as mentioned in the introduction,
\citet{mayor11} point to a minimum in the planet distribution at $\sim 40\mearth$ as being
consistent with the \citet{mordasini09a,mordasini09b} prediction of a ``moderate" planet desert in the range
30--$100\mearth$, with a reduction of a factor of 2--3 from the number of Jupiter-mass planets.
Our analysis does indicate a minimum at $\sim 30\mearth$ in the $3.75 < P < 30\,$days bin, but
our error bar estimates indicate that this is not significant even at the 1--$\sigma$ level. It is 
possible that this minimum might appear more significant with a different binning, but
we will address this issue in Section~\ref{sec:forward} with a forward modeling approach that
involves no binning. Also, note that there is no evidence for the predicted factor of 2--3 deficit 
of planets at 30--$100\mearth$ compared to Jupiter-mass ($\sim 318\mearth$) planets \citep{mordasini09a,mordasini09b} in
any of the period ranges shown in Figure~\ref{fig:mayor_hist}. There does seem to be a
marginal increase of Jupiter-mass planets compared to 100--$200\mearth$ planets in the
240--$1920\,$days period range, but this is only marginally significant and is not duplicated
at shorter or longer periods. 

In this discussion, we have avoided the distinction between planet masses, $M$, and the 
$M\sin i$ values that are measured by radial velocity monitoring. Because small inclinations,
$i \ll 1$, are disfavored geometrically, this is generally expected to have a small effect, but it
could conceivably be important if the occurrence rate, $\eta$, is strongly dependent on planet mass. 
Fortunately, this complication disappears with the forward modeling approach presented in the next
section.

\section{Forward Modeling Analysis} \label{sec:forward}

We now turn to a more rigorous forward modeling analysis to search for evidence of this predicted desert,
attributed to the runaway gas accretion process of the core accretion theory. Forward modeling allows us to 
investigate the possibility of a desert with somewhat different parameters than proposed by 
\citet{idalin04} and \citet{mordasini09a,mordasini09b}, and it enables us to avoid the uncertainties relating to the binning of data
and error bar estimates mentioned in the previous section. 

\subsection{Exoplanet Desert Occurrence Rate Models}\label{sec:forward-des}

Following earlier Doppler radial velocity analyses \citep{cumming08,fernandes19,wittenmyer20}, we base our 
model on power-laws in both mass, $M$, and period, $P$. The previous analysis of the CORALIE/HARPS
sample \citep{fernandes19} found a break in the planet occurrence rate at a period of $P = 2100\pm 1200\,$days,
which is consistent with the $P = 1920\,$days boundary between the third and fourth period ranges presented in 
Figure~\ref{fig:mayor_hist}. (\citet{fernandes19} indicate that this broken power-law applies to masses 
in the range 30-6000$\,\mearth$, but Figure~\ref{fig:mayor_hist} makes it clear that the \citet{mayor11} sample
has little sensitivity for $M < 100\mearth$ at periods longer than the break.)
Since the $M \sin i $ dependence does not appear to follow the same 
power law in these four period ranges, we elect to fit these four period ranges separately without any
power-law break. 

In order to describe a planetary desert, we add the following desert reduction factor, $D(P,M_{\rm des})$ to the 
power law model:
\begin{equation}
D(M,M_{\rm des}) = 1 - B \exp\left( -{\left[\log M-\log M_{\rm des}\right]^2\over 2\sigma_{\rm des}^2} \right) \ ,
\label{eq-desert}
\end{equation}
where the planet desert is centered at period $M_{\rm des}$ with a Gaussian dispersion of $\sigma_{\rm des}$
in $\log P$. (Throughout this paper, $\log$ refers to the base-10 logarithm.) This desert reduction factor multiplies
the power-law factors to yield our planet desert mass function model:
\begin{align}
f_{\rm des}(M,P;A,B,\log M_{\rm des},\sigma_{\rm des},p,m) &\equiv \frac{d^{2} N_{\rm pl}}{d\,\log M\ d\,\log P} \nonumber \\
&= A \left(\frac{P}{1995\,{\rm days}} \right)^{p} D(M,M_{\rm des}) \left({M\over 100\mearth}\right)^{m} .
\label{eq-pow}
\end{align}
We select the Gaussian function to describe the desert because we think that a smooth function is more plausible
than a step function, and we do find that it provides a reasonably good description of the planet distribution predicted by the
most recent version of the Bern group's planetary population synthesis model 
\citep{emsenhuber_ngpps1,emsenhuber_ngpps2} in some period ranges, as discussed in Section~\ref{sec:bern}.
Note that we have selected the pivot point for the period power-law to be $10^{3.3} = 1995\,$days in order to 
match the break in the power-law when we include all periods longer than $3.75\,$days, as discussed in 
Section~\ref{sec:power}.

One complication for our forward modeling
effort is that the completeness, $C(M\sin i,P)$, depends on the inclination angle, $i$, while intrinsic planet properties 
will depend only on the planet mass, $M$. Fortunately, it is well established that there is no physical process
that is likely to produce a non-random distribution of planetary orbital planes as seen from Earth, so we can 
treat the inclination angles, $i$, as a random variable to integrate over. We introduce the variable $\Msi \equiv M \sin i $
to refer to the observable parameter measured by the Doppler radial velocity method.

\begin{figure}
\begin{center}
\includegraphics[width=150mm]{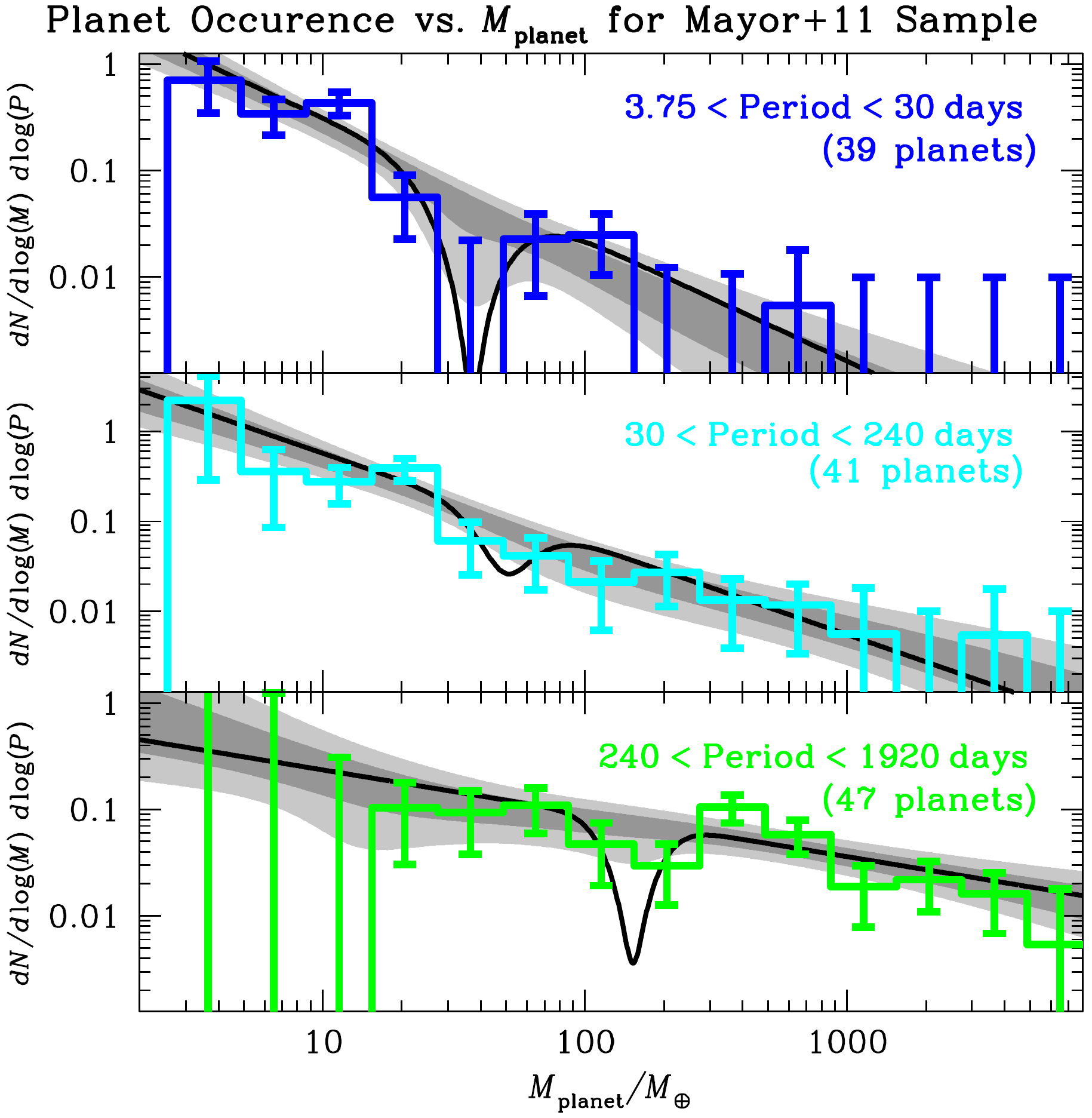} 
\end{center}
\caption{
The desert power-law planet occurrence rate models compared to the histograms from the top three 
period range panels of Figure~\ref{fig:mayor_hist}. The black curves are the best fit models, and the
dark and light grey shaded regions indicate the central 68.3\% and 95.4\% distribution of planet
occurrence rates as a function of mass from the MCMC distributions. These indicate no significant
evidence of a sub-Saturn-mass planet desert in the \citet{mayor11} CORALIE/HARPS data set.
\label{fig:desmod}
}
\end{figure}

We use a standard forward modeling Bayesian likelihood analysis \citep{alcock96,suzuki16} to determine the 
posterior distribution of models that are consistent with the data. The likelihood function is given by
\begin{align}
\mathcal{L}_{\rm des}(&A,B,\log M_{\rm des},\sigma_{\rm des},p,m) = e^{-N_{\rm exp}} \nonumber \\
&\times \prod _{j}^{N_{\star}}\int di\,\sin i\, f_{\rm des}(\Msi_{j}/\sin i,P_{j};A,B,\log M_{\rm des},\sigma_{\rm des},p,m)\, C(\Msi_j,P_j)  \ ,
\label{eq-L_des}
\end{align}
where
\begin{equation}
N_{\rm exp} = \iiint dM\, dP \, di\, \sin i\, f_{\rm des}(M,P;A,B,\log M_{\rm des},\sigma_{\rm des},p,m)\, C(M\sin i,P) \ ,
\label{eq-Nexp_des}
\end{equation}
is the expected number planets detected for the model with the given parameters: $A$, $B$ ,$\log M_{\rm des}$,
$\sigma_{\rm des}$, $p$, and $m$. We evaluate this likelihood function with a Markov Chain Monte Carlo (MCMC)
for each of the period ranges: 3.75--$30\,$days, 30--$240\,$days, 240--$1920\,$days and 1920--$15260\,$days,
with chains of 140,000-330,000 steps after 2000 burn-in steps. The results of these calculations are summarized in
Figure~\ref{fig:desmod}. The best fit models are shown as the black curves, while the dark and light grey 
shading indicates the central 68.3\% and 95.4\% likelihood distributions for the planet occurrence rate as
a function of mass from the MCMC calculations. These plots are shown for the median $\log P$ values
for each of the period ranges. These central periods are 
$10.6\,$days for $3.75 < P < 30\,$days,
$84.9\,$days for $30 < P <  240\,$days, and
$679\,$days for $240 < P <  1920\,$days.

The histograms of the exoplanet occurrence from the CORALIE/HARPS data set from the 
top three panels of Figure~\ref{fig:mayor_hist} are also shown in Figure~\ref{fig:desmod}. However, the
$x$-axis for Figure~\ref{fig:desmod} is the planet mass, $M$, instead of the Doppler radial velocity
observable $\Msi = M\sin i$. So, we shift the $x$-axis from $\Msi$ to $M \approx \Msi/\sin i_{\rm med} = 1.1547$,
where $i_{\rm med} = \cos^{-1}(0.5)$ is the median inclination value, assuming a
random orientation of the orbital planes. This has no effect on the Bayesian likelihood analysis, since we 
integrate over all inclination angles, $i$, which uses the measured period and $\Msi$ values
without any binning.
The parameters of the best fit and average desert power-law models are given in Table~\ref{tab:des_param}.
The scatter in the normalization parameter $A$ is large because the pivot point of the period power
law is set to a fairly large period of $1995\,$days, and this results in a relatively large scatter in the 
normalization parameter, $A$. Therefore, we also include a parameter, $A_{P\rm m} = A(P_{\rm m}/1995\,{\rm days})^p$,
where $P_{\rm m}$ logarithmic median of the period range being considered. For the different period ranges, we have
$P_{\rm m} = 10.607$, 84.853, and $678.820\,$days for the 3.75--30, 30--240, and 240--$1920\,$days ranges, respectively.
These $A_{P\rm m}$ values reflect the variation in the normalization in the middle of the period ranges under
consideration.

\begin{table}
\caption{Exoplanet Desert Power-Law Model Parameters}
\label{tab:des_param}
\vspace{-0.6cm}
\begin{center}
\begin{tabular}{lcccccc}
\tableline \tableline
parameter & \multicolumn{6}{c} {Period ranges}  \\
   &  \multicolumn{2}{c} {3.75--30\,days} & \multicolumn{2}{c} {30--240\,days} & \multicolumn{2}{c} {240--1920\,days}  \\
   & best & average & best & average &best & average \\
\tableline
$A$ &                                 0.745 & $4.03\pm 2.44$ & 0.104 & $0.38\pm 0.32$ & 0.105 & $0.178\pm 0.087$ \\ 
$A_{P\rm m}$ &                0.0177 & $0.020\pm 0.009$ & 0.0561 & $0.068\pm 0.022$ & 0.0920 & $0.125\pm 0.052$ \\ 
$B$ &                                 0.977 & $0.52\pm 0.26$ & 0.767 & $0.48\pm 0.24$ & 0.954 & $0.51\pm 0.27$ \\ 
$\log M_{\rm des}$ &          1.529 & $2.08\pm 0.48$ & 1.698 & $2.08\pm 0.42$ & 2.183 & $1.61\pm 0.44$ \\ 
$\sigma_{\rm des}$ &        0.155 & $0.45\pm 0.28$ & 0.137 & $0.55\pm 0.28$ & 0.110 & $0.53\pm 0.29$ \\ 
$p$ &                                 0.714 & $0.97\pm 0.18$ & 0.194 & $0.473\pm 0.23$ & 0.123 & $0.30\pm 0.24$ \\
$m$ &                                $-1.229$ & $-1.18\pm 0.15$ & $-1.008$ & $-0.94\pm 0.12$ & $-0.407$ & $-0.49\pm 0.14$ \\
\tableline
\end{tabular}
\vspace{-2mm}
\end{center}
\end{table}

Figure~\ref{fig:desmod} indicates that the best fit models do have narrow exoplanet deserts, and the 
deserts in the 3.75--$30\,$days and 30--$240\,$days period ranges are centered in the range of 10--$100\,$days
predicted by \citet{idalin04}. The depth of the desert features in the best fit models, represented by the $B$, is
fairly deep at their centers, reducing the occurrence rate by over 95\% for the 3.75--$30\,$day and 240--$1920\,$day ranges and 
by 76.7\%in the 30--$240\,$day range.
However, the distribution of models consistent with the data, as represented by
the grey shaded regions in Figure~\ref{fig:desmod} tell a different story. The central 68.3\% and 95.4\%
model occurrence rate distributions generally do not follow the best fit models, although the lowest
extent of the central 95.4\% occurrence rate distribution for the 3.75--$30\,$days period range does
have a dip at the location of the best fit desert model, but there is virtually no such feature in the 
central 68.3\% occurrence rate distribution. These features are also narrower than the \citet{idalin04} prediction
of a desert spanning a factor of 10 in mass. The Gaussian $\sigma_{\rm des}$ 
values (in $\log M$) range from 
$\sigma_{\rm des} = 0.110$ for the 240--$1920\,$days period range to 
$\sigma_{\rm des} = 0.155$ for the 3.75--$30\,$days period range. These imply full-width half-max
values in the range 1.8--2.3 in $M$.
The most likely explanation for this is that the very narrow desert features 
of the best fit models are due to statistical noise in the observed exoplanet distribution. This would certainly
explain why the central 68.3\% range of the occurrence rate distribution does not show these features.

We do note that the location of the best fit desert in the 3.75--$30\,$day  and 30--$240\,$day period ranges do
roughly correspond to the
minimum at $M \sin i = 40 \mearth$ mentioned by \citet{mayor11}, particularly in their sample with
periods $< 100\,$days. However, our analysis indicates that this minimum is not likely to be significant.
In fact, this conclusion seems consistent with the completeness corrected histogram shown
in Figure~12 of \citet{mayor11}, where the bin centered near $M \sin i = 40 \mearth$ is consistent
at 1--$\sigma$ with the next two larger bins. The large rise at lower masses is significant, as is the case with 
our histogram in the top panel of Figure~\ref{fig:desmod}, although our error bars are more conservative. 
However, this is does not support the runaway accretion prediction that planets should grow rapidly across 
the 10--$100\,\mearth$ mass interval.

\subsection{Power-Law Occurrence Rate Models}\label{sec:power}

\begin{table}
\caption{Pure Power-Law Model Parameters}
\label{tab:pow_param}
\vspace{-0.3cm}
\begin{center}
\begin{tabular}{lcccccc}
\tableline \tableline
parameter & \multicolumn{6}{c} {Period ranges}  \\
   &  \multicolumn{2}{c} {3.75--30\,days} & \multicolumn{2}{c} {30--240\,days} & \multicolumn{2}{c} {240--1920\,days}  \\
   & best & average & best & average &best & average \\
\tableline
$A$ &                                 0.542 & $1.77\pm 0.62$ & 0.081 & $0.22\pm 0.16$ & 0.099 & $0.113\pm 0.035$ \\ 
$A_{P\rm m}$ &                0.0148 & $0.0148\pm 0.0044$ & 0.0481 & $0.048\pm 0.009$ & 0.0812 & $0.083\pm 0.016$ \\ 
$p$ &                                 0.701 & $0.90\pm 0.12$ & 0.165 & $0.40\pm 0.23$ & 0.179 & $0.26\pm 0.24$ \\
$m$ &                                $-1.251$ & $-1.26\pm 0.13$ & $-0.972$ & $-0.98\pm 0.11$ & $-0.367$ & $-0.38\pm 0.09$ \\
\tableline
\end{tabular}
\vspace{-2mm}
\end{center}
\end{table}

Since we have concluded that there is no significant indication of a planet desert in the 
CORALIE/HARPS data set, it is sensible to consider simpler models that do not include the
desert. The model parameters for such models are presented in Table~\ref{tab:pow_param}, which 
includes both the best fit models and the MCMC averages. These power-law models are given
by equation~\ref{eq-pow} with $B\equiv 0$ and therefore $D(M,M_{\rm des}) \equiv 1$.
As was the case for the planet desert models,
the parameter $A_{P\rm m}$ is not a model parameter. Instead, it reflects the normalization of the
occurrence rate function at the logarithmic center of the period range under consideration. So, the scatter in 
$A_{P\rm m}$ is a better indication of the uncertainty in the occurrence rate at the center of the period
range under consideration than $A$. 

We can compare these power-law models to the corresponding power-law plus desert models to see if
the addition of the desert feature significantly improves the fit. We can do this by comparing the 
$\log {\cal L }$ values for the models without and with the desert features. We find that the 
desert features improve the $\log {\cal L }$ values by 1.526, 0.722, and 0.944 for the
3.75--30, 30--240, and 240--$1920\,$days period ranges, respectively. Using frequentist statistics,
these would imply that the pure power-law models were disfavored by $p = 0.030$, 0.189, and 0.115,
for these three period ranges. Following equation 27 of \citet{trotta08}, we can compute upper bounds on the 
Bayes factor of 3.52, 1.17, and 1.48 for the 3.75--30, 30--240, and 240--$1920\,$days period ranges.
These correspond to weak evidence favoring the power-law plus desert model for the 
3.75--$30\,$days period range, and no significant evidence in favor of the desert additions to the 
power-law models at the longer period ranges. While there may be marginal evidence that 
the desert model is an improvement over the pure power-law model for the 3.75--$30\,$days
period range, we do not consider this to be evidence in favor of the desert feature itself based
upon the MCMC results.

\begin{table}
\caption{Broken Power-Law Model Parameters for $0.1 M_{\rm Jup} \leq M\sin i \leq 20 M_{\rm Jup}$}
\label{tab:bpl_param}
\vspace{-0.2cm}
\begin{center}
\begin{tabular}{lccc}
\tableline \tableline
parameter & \multicolumn{2}{c} {$P$ range: 1--11000\,days}  \\
   & best & average & EPOS \\
\tableline
$A$ &                                 0.167  & $0.165\pm 0.035$  & $0.157^{+0.036}_{-0.030}$\\  
$\log(P_{\rm br})$ &            3.32  & $3.20\pm 0.17$ & $3.32^{+0.19}_{-0.24}$ \\
$p_1$ &                             0.553  & $0.592\pm 0.085$ & $0.70^{+0.32}_{-0.16}$ \\
$p_2$ &                         $-0.75$ & $-0.99\pm 0.53$ & $-1.20^{+0.92}_{-1.26}$ \\
$m$ &                            $-0.462$ & $-0.475\pm 0.075$ & $-0.46\pm 0.06$ \\
\tableline
\end{tabular}
\vspace{-2mm}
\end{center}
\end{table}

Finally, to compare with \citet{fernandes19}, we also consider a model with a broken power law in
the period with masses constrained to the range $0.1 M_{\rm Jup} \leq M\sin i \leq 20 M_{\rm Jup}$.
This mass range is considered \citep{fernandes19} to be the range of giant planets.
Our broken power-law mass function model is given by 
\begin{align}
f_{\rm bpl}(M,P;A,P_{\rm br},&p_1,p_2,m) \equiv \frac{d^{2} N_{\rm pl}}{d\,\log M\ d\,\log P} \nonumber \\
&= A \left[\left(\frac{P}{P_{\rm br}} \right)^{p_1}\Theta(P-P_{\rm br}) +  \left(\frac{P}{P_{\rm br}} \right)^{p_2}\Theta(P_{\rm br}-P)\right] M^{m},
\label{eq-brpow}
\end{align}
where $\Theta$ is the Heavyside theta function (or step function), $P_{\rm br}$ is the period break and 
$p_1$, $p_2$, and $m$ are the power-law exponents for small periods, large periods and the planet mass,
$M$, respectively. The likelihood function then becomes
\begin{align}
\mathcal{L}_{\rm bpl}(A,P_{\rm br},&p_1,p_2,m) = e^{-N_{\rm exp}} \nonumber \\
&\times \prod _{j}^{N_{\star}}\int di\,\sin i\, f_{\rm bpl}(\Msi_{j}/\sin i,P_{j};A,P_{\rm br},p_1,p_2,m)\, C(\Msi_j,P_j)  \ ,
\label{eq-L_bpl}
\end{align}
and the number of expected planet detections becomes
\begin{equation}
N_{\rm exp} = \iiint dM\, dP \, di\, \sin i\, f_{\rm bpl}(M,P;A,P_{\rm br},p_1,p_2,m)\, C(M\sin i,P) \ .
\label{eq-Nexp_bpl}
\end{equation}
Evaluating this with a MCMC, as we did with the planet desert models, we find the best fit model, as wll
the range of models from the MCMC distribution. The model parameter results are shown in the first three
columns of Table~\ref{tab:bpl_param} indicating a power law break at 
$P_{\rm br} = 10^{3.20\pm 0.17} = 1600^{+800}_{-500}\,$days. Note that we have used $\log (P_{\rm br})$
instead of $P_{\rm br} $ as the parameter to indicate the location of the break in the period distribution
in our MCMC calculations. This is 
equivalent to assuming that the prior distribution of the $P_{\rm br} $ variable in uniform in $\log (P_{\rm br})$ rather
than uniform in $P_{\rm br}$. These results are consistent with the \citet{fernandes19} results for the same
model as determined by the EPOS program, which are reproduced in the fourth column of Table~\ref{tab:bpl_param}.
(Note that the normalization parameter, $A$, has been modified from the value in the \citet{fernandes19} paper to account
for the fact that they define the broken power law mass function over $d\,\ln M\ d\,\ln P$, whereas we use
$d\,\log_{10} M\ d\,\log_{10} P$.)
Thus, we confirm the main conclusion of \citet{fernandes19} that the giant planet occurrence rate has 
a maximum in the vicinity of the snow line.

\section{Comparison of CORALIE/HARPS Data and Gaussian Desert Model to Generation III Bern Models}
\label{sec:bern}

\begin{figure}
\begin{center}
\includegraphics[width=74mm]{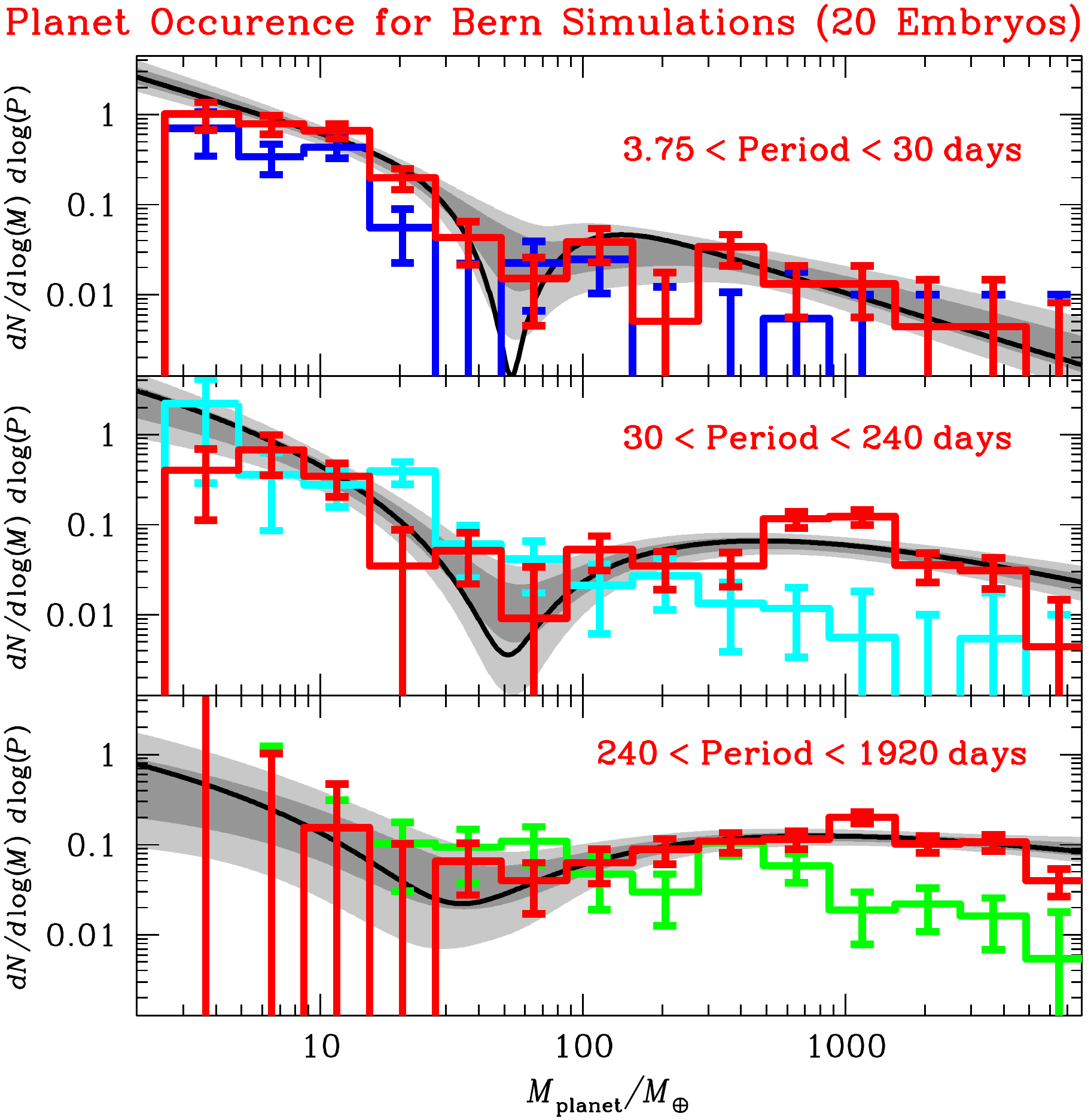} 
\includegraphics[width=75mm]{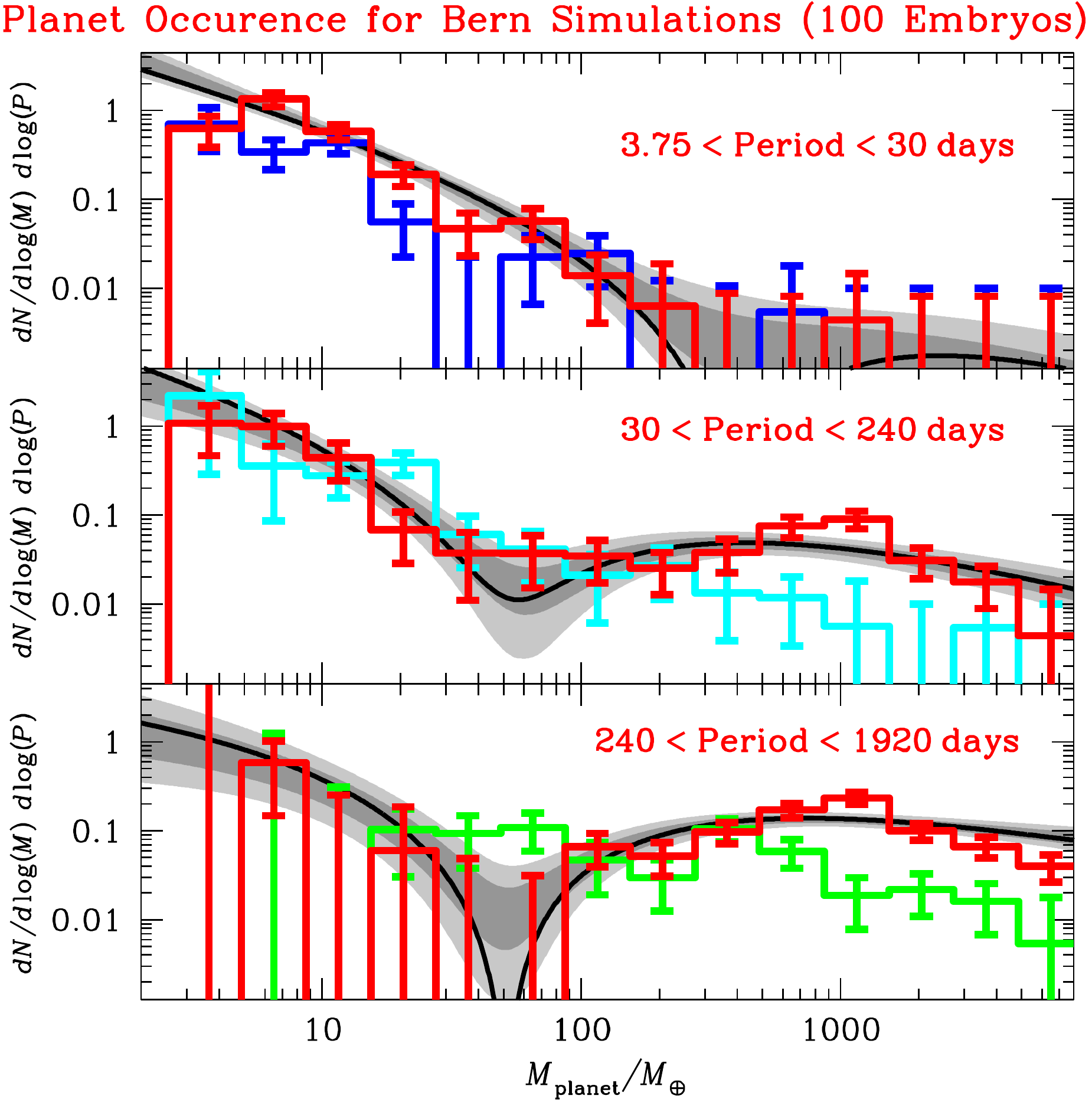} 
\end{center}
\caption{
Comparison of Bern New Generation Planetary Population Synthesis 
calculations (red histograms) to \citet{mayor11} data (blue,
cyan and green histograms) and our Gaussian desert model (black curves and grey shaded regions) using
the same three period ranges used in our previous analyses: 3.75--30\,days, 30--240\,days, and 240-1920\,days.
Results from simulations starting with 20 $0.01\mearth$ planetary embryos are shown in the left panel and 
results from 100 $0.01\mearth$ planetary embryo simulations are shown on the right.
\label{fig:bern}
}
\end{figure}

In this section, we compare the CORALIE/HARPS exoplanet sample \citep{mayor11} to the results
Generation III Bern population synthesis model of planetary formation and evolution 
\citep{emsenhuber_ngpps1,emsenhuber_ngpps2}, and we also fit these model results with our
Gaussian desert, power-law mass function model for the three period ranges used in this paper:
3.75--30\,days, 30--240\,days, and 240-1920\,days. \citet{emsenhuber_ngpps2} present results for different
numbers of $0.01\mearth$ planetary embryos in the initial conditions, and they consider their runs with
100 planetary embryos to be their ``nominal" or most realistic results. However, they also indicate that their
runs with 10, 20, and 50 planetary embryos yield results that are similar to the 100 embryo results, and 
\citet{fernandes19} compared the \citet{mayor11} CORALIE/HARPS sample to an early run of the 
\citet{emsenhuber_ngpps2} simulations with 20 planetary embryos. So, we have presented comparisons to
both the \citet{emsenhuber_ngpps2} 20 and 100 embryo simulations in Figure~\ref{fig:bern}. The
\citet{emsenhuber_ngpps1,emsenhuber_ngpps2} simulations consider only solar mass host stars, but this is
similar to the \citet{mayor11} sample which is dominated by solar type stars in the mass range 
$0.7\msun \leq M_{\rm host} \leq 1.3\msun$.
The Gaussian desert model parameters for these models are given in Table~\ref{tab:des_param_bern}.
Each \citet{emsenhuber_ngpps2}  simulation includes 1000 simulated host stars, so it is comparable to
the 822 stars in the \citet{mayor11} CORALIE/HARPS sample. Our analysis of the simulated planet sample is done in
exactly the same way as our analysis of the CORALIE/HARPS sample, with orbital inclinations assigned randomly.

\begin{table}
\caption{Exoplanet Desert Power-Law Model Parameters for New Bern Models}
\label{tab:des_param_bern}
\vspace{-0.3cm}
\begin{center}
\begin{tabular}{lcccccc}
\tableline \tableline
& \multicolumn{6}{c} {Period Ranges for 20 Embryos}  \\
parameter    &  \multicolumn{2}{c} {3.75--30\,days} & \multicolumn{2}{c} {30--240\,days} & \multicolumn{2}{c} {240--1920\,days}  \\
   & best & average & best & average &best & average \\
\tableline
$A$ &                                 0.04216 & $0.16\pm 0.12$ & 6.644 & $11.8\pm 5.4$ & 0.1789 & $0.132\pm 0.058$ \\ 
$A_{P\rm m}$ &                0.08073 & $0.097\pm 0.024$ & 0.3563 & $0.302\pm 0.075$ & 0.3754 & $0.255\pm 0.098$ \\ 
$B$ &                                 0.9913 & $0.84\pm 0.10$ & 0.9934 & $0.970\pm 0.026$ & 0.9591 & $0.87\pm 0.011$ \\ 
$\log M_{\rm des}$ &          1.726 & $1.83\pm 0.13$ & 1.707 & $1.726\pm 0.070$ & 1.509 & $1.55\pm 0.18$ \\ 
$\sigma_{\rm des}$ &        0.245 & $0.41\pm 0.17$ & 0.850 & $0.80\pm 0.12$ & 0.960 & $0.74\pm 0.17$ \\ 
$p$ &                               $0.124$ & $0.03\pm 0.17$ & 0.927 & $1.12\pm 0.19$ & $-0.687$ & $-0.63\pm 0.14$ \\
$m$ &                                $-0.887$ & $-0.859\pm 0.067$ & $-0.621$ & $-0.583\pm 0.073$ & $-0.334$ & $-0.22\pm 0.11$ \\
\tableline \tableline
& \multicolumn{6}{c} {Period Ranges for 100 Embryos}  \\
parameter    &  \multicolumn{2}{c} {3.75--30\,days} & \multicolumn{2}{c} {30--240\,days} & \multicolumn{2}{c} {240--1920\,days}  \\
   & best & average & best & average &best & average \\
\tableline
$A$ &                                 0.00580 & $0.96\pm 0.65$ & 12.86 & $8.9\pm 2.3$ & 0.3044 & $0.28\pm 0.10$ \\ 
$A_{P\rm m}$ &                0.0877 & $0.074\pm 0.028$ & 0.3941 & $0.274\pm 0.066$ & 0.4348 & $0.39\pm 0.12$ \\ 
$B$ &                                 0.9993 & $0.84\pm 0.17$ & 0.9820 & $0.946\pm 0.028$ & 0.9996 & $0.97\pm 0.028$ \\ 
$\log M_{\rm des}$ &          2.608 & $2.48\pm 0.24$ & 1.727 & $1.758\pm 0.082$ & 1.708 & $1.696\pm 0.082$ \\ 
$\sigma_{\rm des}$ &        0.848& $0.75\pm 0.20$ & 1.104 & $0.80\pm 0.12$ & 0.757 & $0.74\pm 0.14$ \\ 
$p$ &                                $-0.081$ & $0.42\pm 0.20$ & 1.104 & $1.10\pm 0.11$ & $-0.331$ & $-0.34\pm 0.14$ \\
$m$ &                                $-0.896$ & $-0.918\pm 0.087$ & $-0.733$ & $-0.661\pm 0.078$ & $-0.388$ & $-0.32\pm 0.10$ \\
\tableline
\end{tabular}
\vspace{-2mm}
\end{center}
\end{table}

In contrast to the case with the CORALIE/HARPS sample, the \citet{emsenhuber_ngpps2} simulations show
very significant desert features for all period ranges with maximum depths ($B$ values) ranging from 
$0.838 \pm 0.168$ to $0.974 \pm 0.028$. The widths of these features (in $\log M$) are also larger with 
$\sigma_{\rm des}$ values ranging from $0.414\pm 0.167$ to $0.804\pm 0.116$. The one peculiar
result is for the 100 embryo simulations in the 3.75--30\,day period range, where center of the
desert occurs above $100\mearth$ at $\log M_{\rm des} = 2.475\pm 0.236$, but this is largely 
because this data set has only 2 of the 98 simulated planets with $M\sin i > 100\mearth$.

Since the \citet{mayor11} CORALIE/HARPS sample shows no evidence for a planet desert, it is
not surprising that the population synthesis results do not match the CORALIE/HARPS data. The
simulations indicate the desert features that are not seen in the data, and for the 
30-240\,day and 240--1920\,day period ranges the simulations predict a large excess of 
$\sim 1000\mearth$ planets that are not seen in the observed sample. These discrepancies were
also noted by \citet{emsenhuber_v_mayor}. However, they did their comparison using
observed and simulated histograms of detected planets instead of the detection efficiency
corrected histograms that we present here. The problem with their approach is that 
a steep rise in planet occurrence at $M\sin i \simlt 30\mearth$ with a flat
distribution above $30\mearth$ can be converted
into an apparent desert due to the lower detection efficiency for 30--100$\,\mearth$ planets
compared to $>100 \mearth$ planets.

\section{Discussion and Conclusions}\label{sec:conclude}

Our analysis has demonstrated that there is no evidence for the predicted sub-Saturn-mass
planetary desert \citep{idalin04} in the \citet{mayor11} radial velocity sample. Previously, the \citet{suzuki16} 
study demonstrated that there was no such planetary desert in the 
exoplanet mass ratio distribution of planets found by microlensing. These results are complementary
because microlensing is very sensitive to low-mass planets \citep{bennett96} orbiting beyond the snow line
\citep{gouldloeb92}, while the
radial velocity method is currently not sensitive to planets near the lower mass limit of the predicted
desert, beyond the snow line. However, the sensitivity of the radial velocity method improves for shorter period orbits,
and the sensitivity of the \citet{mayor11} sample extends below the predicted planet desert lower
mass limit of $\sim 10\mearth$ for orbital periods in our first and second period bins with 
$P < 240\,$days. In our third period bin, at $240\,{\rm days}<P<1920\,$days, 
the sensitivity extends down to only $\sim 12\mearth$, but this period range overlaps with the
sensitivity of the microlensing method. So, the combination of the
radial velocity analysis presented here and the earlier microlensing work \citep{suzuki16}
appears to rule out this predicted planet desert over a wide range of planetary
orbits, out to $\sim 10\,$AU.

While the population synthesis models are often referred to as theoretical models, it might be more
reasonable to consider them to be methods that attempt to interpret observational constraints in terms
of the many complicated physical processes that are thought to be involved in the formation of 
planets. As mentioned in the introduction, \citet{suzuki18} have presented a number of possible modifications to the 
\citet{idalin04,idalin04ii,idalin05iii,idalin08iv,idalin08v} and Bern group \citep{mordasini09a,mordasini09b,mordasini15}.
These include a number of processes that could slow gas accretion or terminate it well before the planet
reaches $100\,\mearth$. These include heating of the gaseous envelope by the accretion of planetesimals,
a low disk viscosity, low disk scale height, or early formation of a circumplanetary disk. Runaway growth
is often thought to terminate at lower masses at wider orbits, so gravitational interactions between the
planets could transport lower mass planets to Jupiter-like orbits from the wider orbits where they formed.

Another possibility is that individual planetary systems might form in protoplanetary disks that allow
for runaway gas accretion growth that disfavors an intermediate final planetary mass range that could
be considered a desert. However, if the properties of these disks have a significant variation, then it could
be that these disfavored planetary mass ranges do not line up to create a desert in the combined
mass or mass ratio distribution of a large sample of planetary systems. 
So, it could be that the planet desert feature is simply smoothed out
by large variations in the properties of protoplanetary disks. A wide variation in assumed planet accretion
rates plays a role in theoretical framework developed by \citet{adams21}. This approach is more
phenomenological than the population synthesis approach, and does not seek to describe most of
the detailed physical processes involved in the formation of planets. One of the basic assumptions
of the  \citet{adams21} is that the exoplanet mass function has an approximate power-law form that appears to
describe the planet distribution above $\sim 10\mearth$ according to studies using
a variety of different methods \citep{cumming08,suzuki16,schlaufman18,nielsen19,wagner19}.
\citet{adams21} find that approximately power-law mass functions can be achieved through a combination
of exponential decay rate for protoplanetary disk gas and a gas accretion rate roughly proportional to planet
mass if they also include a random gas accretion efficiency factor that can vary significantly between
different systems.

One issue that has yet to be explored in much detail is the host mass dependence of the wide orbit
planet distribution. It is possible that the predicted planet desert does exist for some particular range
of host masses, although the results presented here tend to disfavor this idea since the \citet{mayor11}
sample consists of FGK stars, while the \citet{suzuki16} sample extends to lower mass host stars.
Nevertheless, Kepler observations have established that planets in short period
orbits have a higher occurrence rate around low-mass stars than around solar type stars
\citep{mulders15}, so it would not be surprising to find a host mass dependence of the
wide orbit planet occurrence rate. However, most of the existing demographic
studies do not allow this. The exiting microlensing studies use only the planet-star mass ratio
\citep{gould10,cassan12,suzuki16}, and the largest of the radial velocity studies \citep{mayor11}
does not provide enough information for such a study. The recent radial velocity analyses of
\citet{wittenmyer20} and \citep{fulton21} do provide the individual detection efficiencies for both the stars with
and without detected planets, which is what is needed. However the \citet{wittenmyer20} study is smaller, and 
both surveys are not as sensitive to low planet masses as the \citet{mayor11} study. 
So, this data set is less suitable for a study of the predicted
sub-Saturn-mass desert for wide orbit planets.

While most of the parameters of planetary microlensing events can only be determined with data taken 
during the event, it is possible to determine the masses of the host stars and planets and their
separation with high angular resolution follow-up observations from the {\it Hubble} Space Telescope
\citep{bennett15,aparna17} or adaptive optics (AO) observations on 8--10m class telescopes. The Keck telescopes
have been the most successful at this, due an AO system that is more effective with faint guide stars
\citep{batista15,bea16,aparna18,aparna20,bennett20,vandorou20,terry21}. A good AO correction is important 
for the interpretation of these follow-up observations because it is generally necessary to confirm the
identification of the host star by confirming the relative proper motion between the candidate host star
and the background source star \citep{koshimoto20}. The MOA Collaboration is in the process of extending
the \citet{suzuki16} statistical sample, while the Keck and {\it Hubble} follow-up program is expanding
with observations and host and planet mass determinations for many more microlens planetary systems
in this extended statistical sample. New radial velocity results have recently been reported from the
California Legacy Survey \citep{fulton21}.
So, we expect significant developments in the understanding of wide orbit exoplanet demographics in 
the next few years.
Ultimately, the {\it Nancy Grace Roman Telescope's} Galactic Exoplanet survey \citep{bennett02,penny18} will 
perform a much more comprehensive statistical study of the occurrence rate of wide orbit planets  at
semi-major axes of $a \simgt 0.5\,$AU
with masses down to the mass of Mars ($0.1\,\mearth$).

\acknowledgments

We thank Fred Adams, B.\ Scott Gaudi, Gijs Mulders, and Ilaria Pascucci for helpful suggestions
and comments on a draft version of this paper.
DPB and CR  were supported by NASA through grant NASA-80NSSC18K0274 and award
number 80GSFC17M0002.
CR is supported by the ANR project COLD-WORLDS of the French \emph{Agence Nationale de la Recherche} 
with the reference ANR-18-CE31-0002.
The results reported herein have benefited from collaborations and/or information exchange within NASA's Nexus 
for Exoplanet System Science (NExSS) research coordination network sponsored by NASA's Science Mission Directorate.




\appendix
\section{Planet Deserts in Kepler Data?}\label{sec:kepler}

There have been extensive discussions in the literature of planet
deserts seen in Kepler data. As mentioned in the introduction, there is the
``hot Neptune desert", also known as the  the ``sub-Jovian desert"
\citep{szabo11,lundkvist16,mazeh16,owen19}. This was an observational discovery, based primarily on Kepler data,
and it is thought to be due to heating of the atmospheres of planets that come very close to their host stars.
There is also the Fulton gap
\citep{fulton17}, which is sometimes referred to as the "sub-Neptune" desert and is 
thought to be caused by photo-evaporation. These deserts occur for only very short period
orbits and are thought to be caused by heating or photoevaporation due to the close 
proximity of the host star, so they are not relevant to the desert that the topic of this paper,
which is predicted by the runaway gas accretion scenario at much wider orbits.

There has also been some, mostly informal, suggestions that the Kepler data shows a sub-Jupiter radius
desert for cooler planets at wider separations. For example, this shows up, somewhat weakly, in Figure~7 of 
\citet{kepler2018} at orbital periods $\simgt 20\,$days, and it also can be seen in Figure~3 of 
\citet{berger20} at an incident flux of $\simlt 10\times$ that received by the Earth. In this section,
we show that this sub-Jupiter radius gap is likely to be real, but that it does not imply a sub-Jupiter
mass gap. 

\begin{figure}
\begin{center}
\includegraphics[width=150mm]{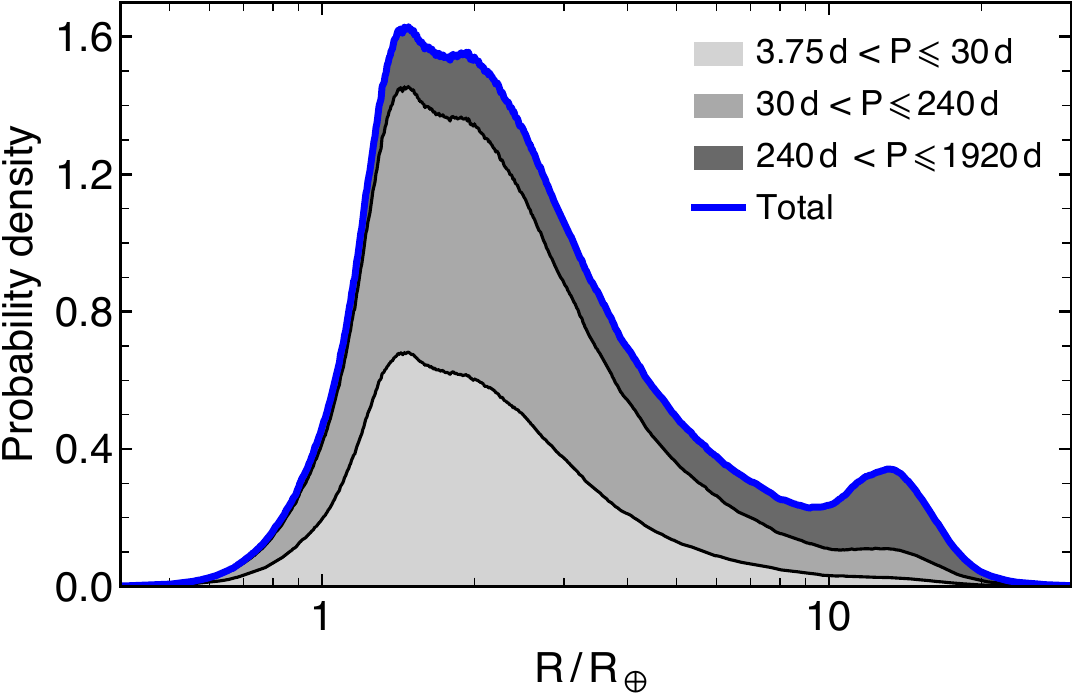} 
\end{center}
\caption{
The predicted radius distribution for planets in the three period bins shown in Figure~\ref{fig:desmod} assuming
the best fit power-law models with no desert feature. The small planet desert feature at a radius of $R \simeq 9 R_\oplus$
is due to the mass-radius relation that reflects the pile-up of planets at radii of $\sim 14\,R_\oplus$ due to the
dominance of electron degeneracy pressure for objects in the mass range of $0.4M_{\rm Jup}$ to $0.08\msun$,
a factor of $\simgt 200$ in mass.
\label{fig:radii}
}
\end{figure}

In order to relate the masses and radii of planets, we use the mass-radius relation of 
\citet{chen17}, and to describe the mass distribution of exoplanets we employ the best
fit pure power-law models from Table~\ref{tab:pow_param} for our three different period
ranges: 3.75--30\,days, 30--240\,days and 240-1920\,days. The results of this calculation
are shown in Table~\ref{fig:radii}. This shows a clear minimum in the planetary 
radius distribution at $R \simeq 9 R_\oplus$ and a maximum at $R \simeq 14 R_\oplus$.
This feature is dominated by planets in the longest period bin of $240\,{\rm days} < P \leq 1920\,{\rm days}$
because the assumed long period mass function has the shallowest mass dependence:
$M^m$ with $m = -0.367$. Note that this exercise is intended to show how a smooth mass
function can still imply a ``planet radius desert", and we do not intend to claim that these
exoplanet mass functions can provide a good description of the Kepler data with its much 
larger exoplanet sample.

The reason that the smooth, monotonic mass function is converted into a radius function 
with a dip and a peak is clear from Figure~3 of \citet{chen17}, which displays their 
mass-radius relation. This relation is nearly flat between masses of $0.4M_{\rm Jup}$ to $0.08\msun$,
a factor of 209 in mass. This is a well known feature caused by the fact that gas giant planets,
brown dwarfs and the lowest mass stars are all largely supported by electron degeneracy pressure
(which is also responsible for the decreasing radius with increasing mass for white dwarfs).
Thus, the observed potential sub-Jupiter desert in the Kepler planetary radius distribution 
is likely to be explained by the mass-radius relation rather than a feature in the exoplanet mass
distribution, since the feature appears for a model that has no desert in the mass distribution.



\end{document}